# Manipulating Coherent Light Matter Interaction: Continuous Transition between Strong Coupling and Weak Coupling in MoS$_2$ Monolayer Coupled with Plasmonic Nanocavities

*Songyan Hou, Landobasa Y. M. Tobing, Xingli Wang, Zhenwei Xie, Junhong Yu, Jin Zhou, Daohua Zhang, Cuong Dang, Philippe Coquet, Beng Kang Tay, Muhammad Danang Birowosuto\*, Edwin Hang Tong Teo\*, Hong Wang\**

S. Hou, Dr. X. Wang, Dr. Z. Xie, Prof. C. Dang, Prof. B. K. Tay, Prof. E. H. T. Teo, Dr. Muhammad Danang Birowosuto, Prof. H. Wang

School of Electrical and Electronic Engineering, Nanyang Technological University, 50 Nanyang Avenue, 639798, Singapore

CINTRA UMI CNRS/NTU/THALES 3288, Research Techno Plaza, 50 Nanyang Drive, Border X Block, Level 6, 637553, Singapore

E-mail: mbirowosuto@ntu.edu.sg; htteo@ntu.edu.sg; ewanghong@ntu.edu.sg

Dr. L. Y. M. Tobing, J. Yu, J. Zhou, Prof. D. Zhang

School of Electrical and Electronic Engineering, Nanyang Technological University, 50 Nanyang Avenue, 639798, Singapore

Prof. Philippe Coquet

CINTRA UMI CNRS/NTU/THALES 3288, Research Techno Plaza, 50 Nanyang Drive, Border X Block, Level 6, 637553, Singapore

Institute of Electronics, Microelectronics and Nanotechnologies (IEMN), CNRS UMR 8520-University of Lille, 59650 Villeneuve d'Ascq Cedex, France

Strong interactions between surface plasmons in ultra-compact nanocavities and excitons in two dimensional materials have attracted wide interests for its prospective realization of polariton devices at room temperature. Here, a continuous transition from weak coupling to strong coupling between excitons in MoS$_2$ monolayer and highly localized plasmons in ultra-compact nanoantenna is proposed. The nanoantenna is assembled by a silver nanocube positioned over a gold film and separated by a dielectric spacer layer. A 1570-fold enhancement in the photoluminescence is observed at weak coupling regime in hybrid nanocavities with thick spacer layers. The interaction between excitons and plasmons is then directly prompted to strong coupling regime by shrinking down the thickness of spacer layer. Room temperature formation of polaritons with Rabi splitting up to 190 meV is observed with a fair polariton loss around 165 meV. Numerical calculations quantify the relation between coupling strength, local density of states and spacer thickness, and reveal the transition between weak coupling and strong coupling in nanocavities. The findings in this work offer a guideline for feasible designs of plasmon-exciton interaction systems with gap plasmonic cavities.

## 1. Introduction

The strong light matter interaction in solid state systems is a central issue in various fascinating quantum devices, including ultra-low threshold lasers[1-2], ultra-fast workfunction switches[3], quantum information systems[4-5] and phase transition modification[6]. The nature of this coupling depends on the formation of bosonic quasiparticles and the dispersion of hybrid states. In weak coupling regime, the exciton radiation efficiency can be greatly modified by the so-called Purcell effect, resulting in an enhanced photoluminescence (PL)[7-11]. However, when the rate of energy exchange exceeds that of dissipation in hybrid systems, a new hybridization state is produced with a characteristic energy splitting phenomenon known as Rabi splitting in optical spectra[12-14]. To achieve strong light matter interaction, it is essential to ensure that the coupling or energy exchange rate should be faster than other intrinsic dissipation rates. In addition, another demand for plasmonic nanocavities is that the splitting energy should be larger than plasmon damping energy (~ 90 meV) at room temperature[15].

In the past decades, much attention has been directed to light matter interactions between highly confined photons (or plasmons) in nanocavities and excitons in molecules or quantum dots[16-17]. To date, strong coupling has been demonstrated in many systems, such as organic molecules in plasmonic resonators[18], quantum dots in photonic nanocavity[17] and quantum well in Distributed Bragg Reflectors[19]. These platforms integrated with nanocavity enable strong light matter interactions to overcome thermal loss. However, organic molecules and quantum dots are known to suffer photobleaching and oxidization, which limit their implementation in polariton devices.

Transition metal dichalcogenide (TMD) semiconductors have gained much interests in the study of photonic devices due to their unique optical properties[20-23]. Of major interests are their high oscillator strength, large exciton binding energy and uniformity of optical properties within the entire flakes. These special features are constructive for strong coupling in TMD at room temperature, thereby opening a new avenue for both theoretical study and practical optoelectronic applications. Recently, Rabi splitting has been observed in various TMDs coupled with plasmonic nanocavities[23-25]. Due to the inverse square dependence of coupling strength on mode volume $g \propto 1/\sqrt{V}$, it is important to have an optical cavity with ultra-small mode volume to achieve strong coupling[4, 24, 26]. Gap plasmon nanocavities exhibit highly localized electric field within ultra-small volume, which enables fast and coherent energy exchanges between emitters and nanocavities. This compact gap plasmonic nanocavities can be facilely assembled by placing nanoparticles over a smooth gold film, separated by an ultra-thin dielectric layer. $WSe_2$ and $WS_2$ monolayers have recently exhibited Rabi splitting coupled with this type of gap plasmon nanocavities[27-28]. However, the effect of dielectric layer thickness on the transition from weak to strong coupling has yet to be explored.

Here, we demonstrate a plasmonic nanocavity based exciton polariton system with tunable coupling strength and on-demand transition between weak coupling and strong coupling. The ultra-small resonators consist of silver nanocubes over an ultra-smooth gold film (NCoM), separated by a dielectric layer. These NCoM cavities are able to confine the optical modes within an ultra-small gap, with tunable resonance frequencies and coupling strength by adjusting the nanocube size or the dielectric layer thickness. Surprisingly, in strong coupling regime with a thin dielectric layer, the ultra-strong optical mode confinement and ultra-small mode volume can give rise to strong coupling regime with Rabi splitting up to 190 meV. We also show that the light matter interactions can be tuned from strong coupling to the weak coupling regime when the thickness of spacer layer is increased, which results in the observation of a 1570-fold PL intensity enhancement due to the Purcell effect. Our theoretical calculations confirm that such a transition from strong coupling to weak coupling is due to the changes of mode volume, coupling strength and electric field intensity. Further, an optimal gap thickness for strong light matter coupling is deduced from calculations. Based on the novel studies on the transition between weak and strong coupling, we believe this work presents an effective guideline to approach feasible designs of plasmon-exciton interaction systems.

## 2. Results and Discussion

The $MoS_2$ was grown on $SiO_2$/Si substrates by chemical vapor deposition (CVD) methods and then transferred onto an ultra-smooth gold film (see Experimental Section). The monolayer characteristic of $MoS_2$ was confirmed by atomic force microscopy (AFM) and the thickness was determined as ~ 0.77 nm (**Figure 1**a). Typical bright and dark field images of hybrid nanostructures are shown in Figure 1, with $MoS_2$ monolayer in triangular shapes and silver nanocubes in black dots in bright field image (Figure 1a) or colorful points in dark field image (Figure 1b). The gap plasmon nanocavity systems consist of self-assembled silver nanocubes on a gold mirror with CVD grown $MoS_2$ monolayer and a thin $Al_2O_3$ spacer layer embedded into the gap (Figure 1c). Briefly, $MoS_2$ monolayer is transferred onto $Al_2O_3$-coated ultra-smooth gold film (Supporting Information Figure S1), followed by drop-casting silver nanocubes to form NCoM systems (Figure 1c) (see Experimental Section). In these plasmonic gap nanocavities, the ultra-small mode volume can yield high optical field confinement which is sufficient to produce both weak coupling and strong coupling[29]. However, incorporating $MoS_2$ monolayer into gold nanosphere positioned over gold film (NSoM) can only give weak coupling with PL enhancement due to the mismatch between the polarization of dipole movement and plasmonic mode in the gap (Supporting Information Figure S2). The plasmon resonance frequencies of NCoMs can be easily tuned from red to green either by adjusting the thickness of spacer layer or the size of nanocube. As shown in **Figure 2**, the individual NCoM constructs are observed as diffraction limited, bright and colorful point scatters in dark field images under white lamp illumination.

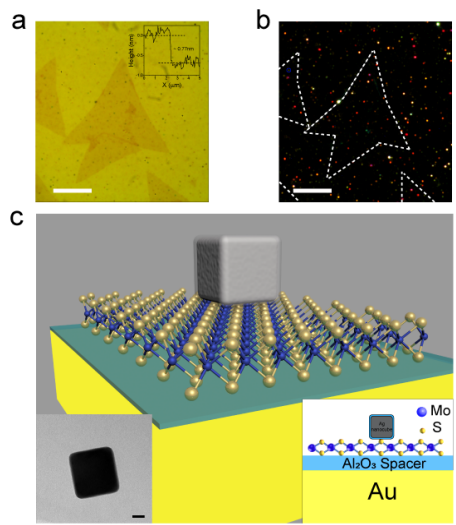

**Figure 1** Construction of NCoM-$MoS_2$ heterostructures. **a)** Bright field of optical image of monolayer $MoS_2$ flake. The inset shows AFM height profile of $MoS_2$ monolayer on $Al_2O_3$ coated gold film. **b)** Dark field scattering image of $MoS_2$ monolayer corresponding to panel a. Scale bar in a and b is 20 µm. **c)** Three-dimensional illustration of a NCoM cavity encapsulating monolayer $MoS_2$ flakes. Insets show a transmission electron microscopy (TEM) image of an individual silver nanocube (scale bar: 20nm) and cross-sectional schematic of a NoCM construct.

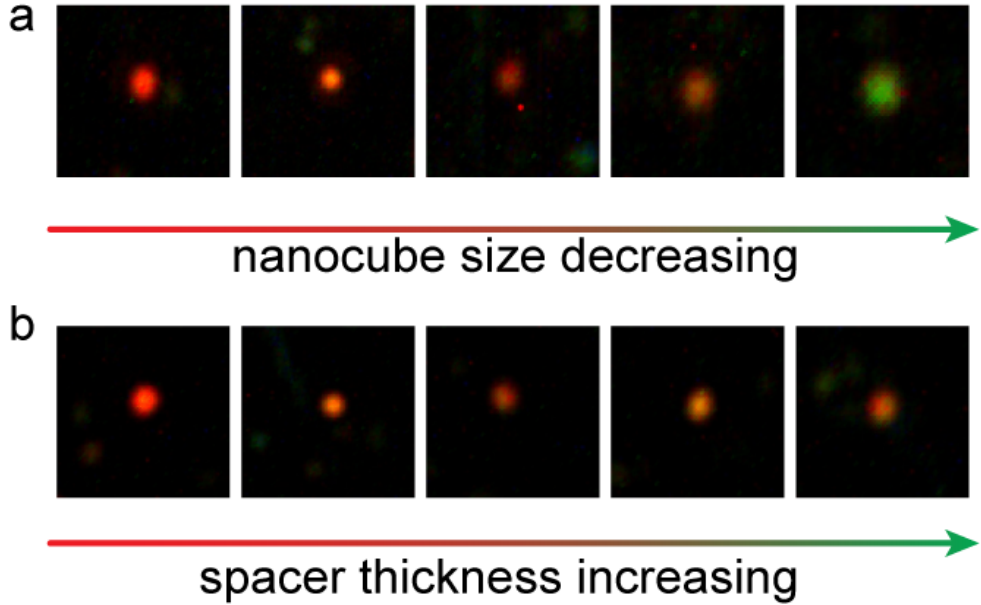

**Figure 2** Tunability of plasmonic cavity resonance in NCoM systems. Tuning cavity resonance **a)** by varying nanocube sizes on a fixed spacer thickness (5 nm) and **b)** by varying spacer thickness with fixed nanocube size (edge length: 80 nm).

MoS$_2$ monolayer is chosen in our study for its large exciton binding energy ($\geq 570$ meV )[30] which results in a stable PL emission (Figure 3a) and allows room temperature strong coupling operation. Typically, there are two excitons named A and B, associated with the splitting of valence band in first Brillouin zone[31]. Usually, only A exciton is manifested as dominant peak in PL spectra due to the low intrinsic quantum efficiency of B exciton[32]. When MoS$_2$ monolayer is incorporated into NCoM with 5 nm spacer on resonance, the strong coupling between excitons and plasmons is observed in PL spectra at bright field (**Figure 3**a). Instead of single A exciton peak on gold film or glass, two split peaks are observed in PL spectra with high energy polariton (HEP) branch at 630 nm and low energy polariton (LEP) branch at 681 nm. Compared to dark field spectra from the same NCoM construct, the splitting energy in PL is smaller than that of dark scattering spectra (Figure 3b). The much smaller splitting energy in PL is attributed to the faster polariton relaxation rate than emission rate and also the difference in splitting between absorption and scattering[33]. Similar to typical strong coupling systems, the HEP emission is much weaker than that of the LEP, which is due to the fast nonradiative rate with low quantum efficiency and the relaxation toward uncoupled excitonic states in HEP branch[19, 33]. Dark field measurements are conducted on many individual NCoM systems with 5 nm $Al_2O_3$ spacer layer in Figure 3c. Tuning plasmon resonance by varying nanocube size results in two split peaks with different spectral contrasts and a characteristic dip at the wavelength of A exciton, indicating the occurrence of strong coupling. The positions of measured nanocubes are labelled in Supporting Information Figure S3, and the same region is then imaged with SEM to determine the corresponding silver nanocube sizes (Supporting Information Figure S4).

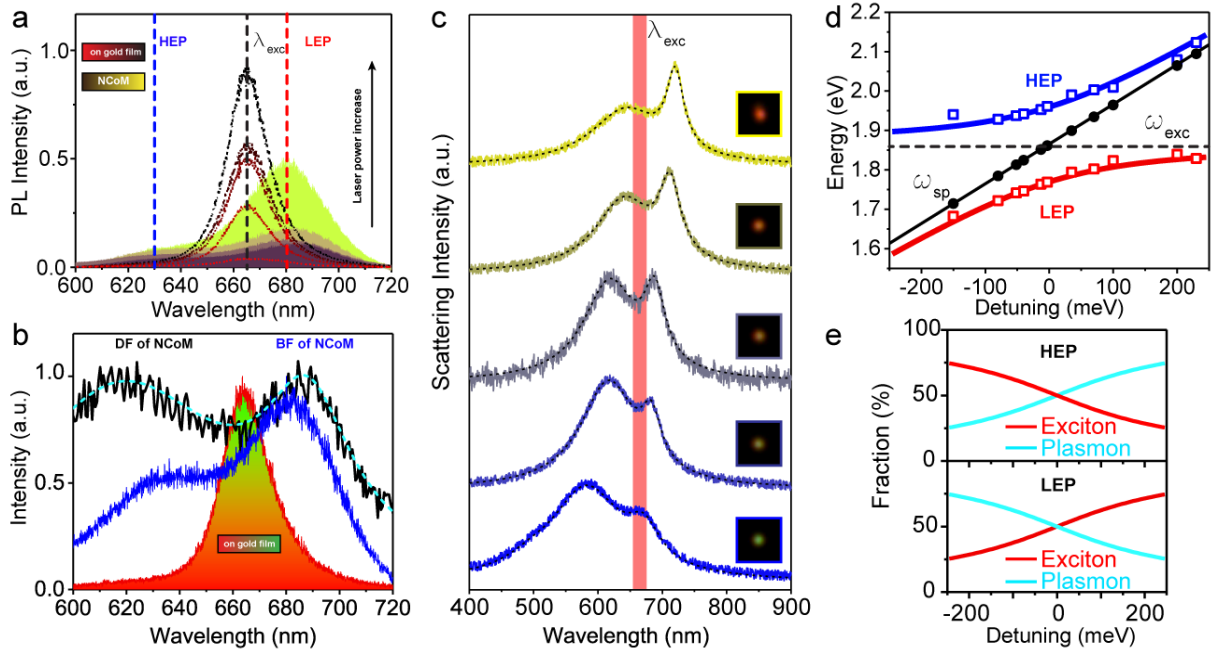

**Figure 3** Signature of strong coupling in NCoM constructs with 5 nm spacer layer. **a)** Power dependent PL emission of $MoS_2$ on gold film (dashed lines) and incorporated into NCoM systems (shadow plots). **b)** Dark field spectra (black) and bright field PL emission (blue) from same NCoM construct. The shadow plot shows the PL spectra from $MoS_2$ monolayer on gold film. **c)** Normalized dark field scattering spectra from different individual NCoMs. Insets are the corresponding scattering images. **d)** High energy polariton (HEP) and low energy polariton (LEP) as a function of detuning. **e)** The contributions of exciton and plasmon to HEP and LEP against detuning.

To describe this plasmon-exciton interaction process, the classical coupled harmonic oscillator model (CHOM) is used to provide an accurate physical analyzer. In this model, the involved excitons are viewed as a "super oscillator", and then the whole process is pictured as two coupled oscillators[17]:

$$\begin{pmatrix} E_{SP} - i\Gamma_{SP}/2 & g \\ g & E_{ex} - i\Gamma_{ex}/2 \end{pmatrix} \begin{pmatrix} \alpha \\ \beta \end{pmatrix} = E_{\pm} \begin{pmatrix} \alpha \\ \beta \end{pmatrix} \quad (1)$$

Here $E_{SP}$ and $E_{ex}$ are the energies of NCoM plasmon and A exciton in MoS$_2$ monolayer, respectively; g is the coupling strength of the hybrid system; $\Gamma_{SP}$ and $\Gamma_{ex}$ represent the line widths of plasmon and A exciton; $\alpha$ and $\beta$ are the eigen coefficients that satisfy $|\alpha|^2 + |\beta|^2 = 1$; and $E_{\pm}$ are the eigen values associated with the energies of the hybrid nanostructure. Given that the widths of plasmon and exciton are small compared to their energies, the energy of the hybrid systems can be estimated as:

$$E_{\pm} = \frac{1}{2}(E_{ex} + E_{SP}) \pm \sqrt{g^2 + \frac{1}{4}(\delta - \frac{i}{2}(\Gamma_{SP} - \Gamma_{ex}))^2} \quad (2)$$

in which $\delta = E_{SP} - E_{ex}$ is the detuning energy between plasmon and A exciton in MoS$_2$ monolayer. The Rabi Splitting $\Omega$ can be calculated at $\delta = E_{SP} - E_{ex} = 0$:

$$\Omega = E_{+} - E_{-} = \sqrt{4g^2 - \frac{1}{4}(\Gamma_{SP} - \Gamma_{ex})^2} \quad (3)$$

The line widths of plasmon and A exciton are extracted to be 280 meV and 50 meV, respectively. Figure 3d shows dispersion curve extracted from the dark field scattering spectra in Figure 3c, together with a fitting from the CHOM model. The results show a clear anti-crossing behavior with a Rabi splitting $\Omega$ = 190 meV at $E_{SP} = E_{ex}$, which satisfies strong coupling criterion ($\Omega > \frac{\Gamma_{SP} + \Gamma_{ex}}{2}$)[34]. The polariton system can be characterized by the comparison of Rabi splitting with polariton loss. People already demonstrated strong coupling using low loss plasmonic structures with narrow resonance width[35-39]. However, this splitting energy is hitherto the largest Rabi energy splitting among all plasmon-exciton polariton systems with TMD 2D materials We also note that the average loss of polariton of 165 meV in our observation is still comparable with that in strong coupling observed in WSe$_2$ and WS$_2$, and better than 100-meV losses that were observed very recently in both mentioned materials[35, 38,39] (see Supporting Information Table S1). The contributions of plasmon and exciton components to the HEP and LEP can also be calculated from fitting (Figure 3e). The plasmon (exciton) constituent gradually dominates HEP (LEP) when the detuning changes from negative to positive. Notably, strong coupling has been demonstrated by coupling MoS$_2$ with Fano resonators[19, 40]. However, owing to the low confinement of optical mode in these nanocavities, the strong coupling can only be clearly observed at cryogenic temperature (77 K) with a small Rabi splitting of 58 meV. Also, a big challenge for Fano resonators lies in the difficulty of resonance tunability. On the other hand, a spectra splitting was also observed in MoS$_2$/Ag nanoparticles hybrid structures[41]. However, only PL enhancement at weak coupling regime was observed without anti-crossing behavior.

By further increasing the spacer thickness to 10 nm and more, only PL enhancement can be observed instead of Rabi splitting. To quantitatively investigate the enhancement performance of NCoMs constructs, a series of experiments were conducted on (i) samples of MoS$_2$ monolayer coupled into NCoMs (ii) samples with MoS$_2$ monolayer transferred on a glass slide without any nanocube. Samples were excited by a 532 nm continuous wave laser and the PL signal was collected by a spectrometer equipped with a photomultiplier tube (PMT) (see Experimental Section). The isolated silver nanocube in NCoMs was marked out from dark field images and subsequently PL measurements were carried out. A tremendous enhancement of ~ 3 times is observed in PL spectra (Figure 4a). Given that the area of PL collection is much larger than that of nanocube which contributes to the PL intensity enhancement, the average enhancement of NCoMs is then quantified by the PL enhancement factor (EF)[15]:

$$EF = \frac{I_{NCoM}}{I_{glass}} \times \frac{A_{spot}}{A_{NC}} \qquad (4)$$

where $I_{NCoM}$ ( $I_{glass}$ ) is the PL intensity from MoS$_2$ monolayer coupled in an individual NCoM (on glass), $A_{spot}$ and $A_{NCoM}$ are the area of laser spot (diameter ~ 1.5 µm) and that of an individual silver nanocube, respectively (Figure 4b inset). The calculated EF from equation (4) is plotted against the thickness of spacer layer in Figure 4b. When the spacer thickness increases from 10 nm to 25 nm, a dramatic decrease in EF is observed from 803 to 225 (Figure 4b). Such a decrease in EF is accompanied by a simultaneous decrease in the Purcell factor (Fp). The calculated Fp (normalized to spacer with 10 nm thickness) exhibits a similar trend to that of EF, indicating that the reduction of EF is attributed to the decline of Fp (see Supporting Information Note5 for details). The enhanced PL and Fp are in good agreement with the faster lifetime in NCoM compared to that on glass (see Supporting Information Figure S5). Although the detuning range of NCoM cavity is large enough reach the energy of B exciton, there is no effect of cavity on B exciton in the spectra. This is attributed to the low quantum efficiency and very high nonradiative loss in B exciton in our MoS$_2$ samples, which are different with other work[42].

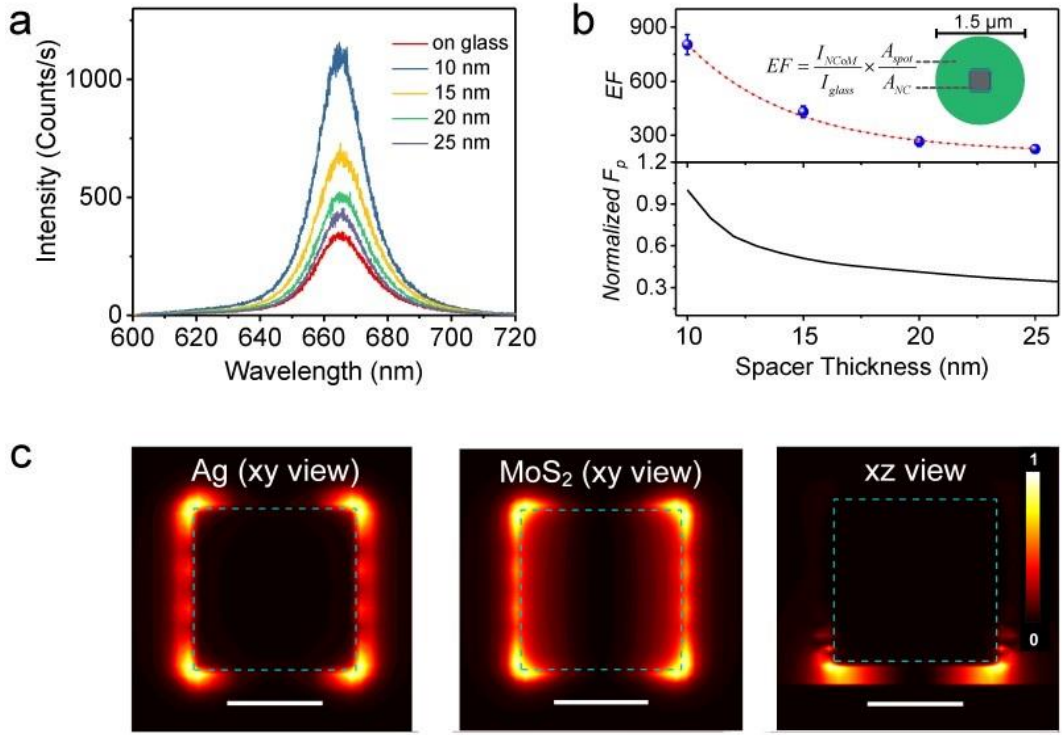

**Figure 4** Photoluminescence enhancement from NCoM constructs at weak coupling regime. **a)** PL spectra from MoS$_2$ monolayer at different environments. **b)** Experimental photoluminescence enhancement factor (upper, with red dashed line to show the guidance) and calculated normalized Purcell factor (lower, normalized to values with 10 nm spacer) as a function of spacer thickness in NCoMs. **c)** Electric field intensity distribution within the nanocube and MoS$_2$ monolayer at $\lambda=665nm$. The position of nanocube is marked with dashed square. Scale bar = 50 nm.

We further investigated the electric field intensity distributions in NCoMs (Figure 4c). We find that the in-plane optical mode is enhanced in the gap at the four corners of the silver nanocube, where it exhibits the largest current density and field enhancement (see Supporting Information Figure S6 and Figure S7). The energy density is mostly distributed in MoS$_2$ monolayer rather than in nanocube due to the large refractive index of MoS$_2$ monolayer (see Supporting Information Figure S8), indicative of higher local density of states (LDOS).

In order to further understand this transition between strong coupling and weak coupling, we calculated the mode volume, coupling strength and electric field intensity in NCoMs with different spacer thicknesses. The coupling strength g is an essential criterion to characterize the strong coupling and can be extracted from equation (3)[16]. Based on the usual definition, g can be estimated as[16]:

$$g = \mu_m \sqrt{\frac{4\pi \hbar N c}{\lambda \varepsilon \varepsilon_0 V}} \tag{5}$$

where $\mu_m$ is exciton transition dipole moment, $N$ is the exciton number, $\lambda$ is the wavelength, $V$ is the mode volume, $\varepsilon$ is the dielectric permittivity and is the speed of light. $N$ can be estimated from the exciton areal density in MoS$_2$, $\sigma_{exc}$, and the mode effective area, $A$. From Equation (5), a large coupling strength can be achieved by reducing the mode volume and optimizing the number of excitons involved in the interaction through the effective area in the emitter. The mode volume can be obtained by the integration of energy density over the volume of NCoM, and normalized to its maximum value, which is still located at the vicinity of the emitter position[18, 43-44]:

$$V(\omega) = \frac{\int W(\vec{r},\omega) d^3 \vec{r}}{\max\left[W(\vec{r},\omega)\right]} \tag{5}$$

where $W(\vec{r},\omega)$ is complex energy density inside the metal and is given by:

$$W(\vec{r},\omega) = \frac{1}{2}\left(\frac{\partial\left[\omega\varepsilon(\vec{r},\omega)\right]}{\partial\omega}\varepsilon_0 \left|E(\vec{r},\omega)\right|^2 + \mu_0 \left|H(\vec{r},\omega)\right|^2\right) \tag{6}$$

where $\varepsilon(\vec{r},\omega)$ is the material permittivity at position $\vec{r}$. In this calculation, a normal incident plane wave is used to illuminate the NCoM and the scattered electric field is collected to represent the plasmonic mode from NCoM. With the thickness of spacer in NCoM decreasing gradually, the mode volume shrinks dramatically owing to the high confinement of plasmon modes while the extracted $g$ increases tremendously (**Figure 5**a) and strong coupling regime occurs when the thickness is smaller than ~ 6 nm. The extracted $g$ in this plot also depends on the number of the excitons through the relationship between the mode effective area, $A$ and the spacer thickness (see Supporting Information Figure S9). We calculated the electric field intensity in different NCoMs, noting that the nanocube size is adjusted to tune the plasmon resonance of NCoM overlapping with A exciton in MoS$_2$ monolayer. We find that the electric field intensity decreases exponentially with the increase of spacer thickness starting from 5 nm, and nonradiative decay channels are dominant due to the absorption from metals when the spacer layer is thinner than 5 nm (Figure 5b, c). Although smaller spacer thickness will enable larger coupling strength, it is faced with both the increase of absorption from metal and the difficulty in the fabrication of homogeneous spacer layer. Based on the reasoning above, embedding MoS$_2$ monolayer into NCoM cavities with thick spacer can trigger weak coupling with high PL enhancement. Shrinking down the spacer layer to decrease mode volume and prompt field confinement can increase the coupling strength, thus switching on light matter interactions at strong coupling regime. It is also clear that NCoM with 5 nm spacer layer offers an optimal platform for strong coupling.

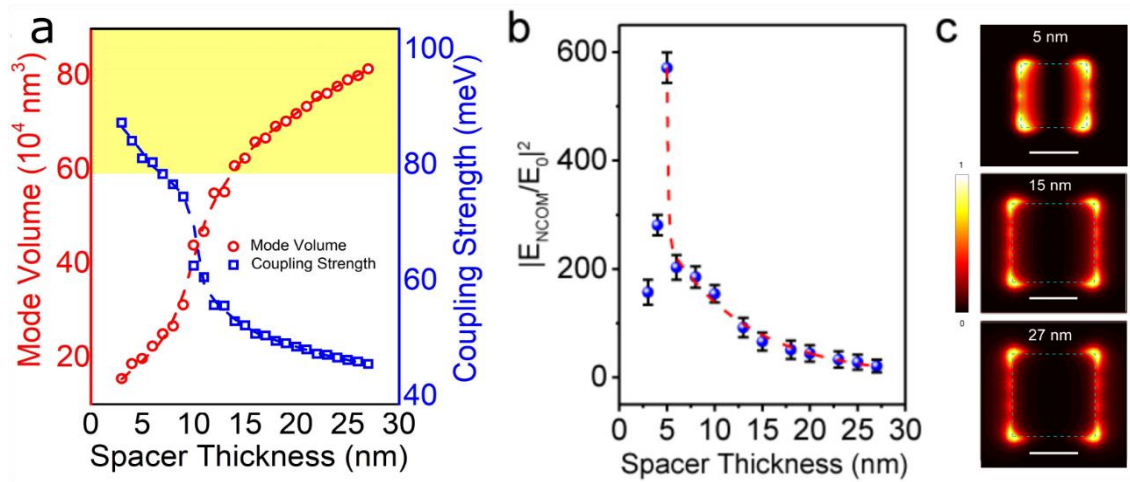

**Figure 5** Comparison of calculated mode volume, coupling strength and electromagnetic energy density in NCoMs with different spacer thickness. **a)** Optical mode volume (red) and coupling strength (blue) vs. the thickness of spacer in NCoM constructs. Strong coupling regime is shaded in yellow. The dashed lines give eye guidance. **b)** Electric field intensity for NCoMs with different spacer thickness. **c)** Maps of electric field intensity for different NCoMs. Noting that the size of nanocube is adjusted to overlap A exciton in MoS$_2$ monolayer. Scale bar: 50 nm.

## 3. Conclusions

The significant role of spacer thickness on strong coupling and weak coupling from MoS$_2$ monolayer in NCoM has been explored experimentally and theoretically. We successfully demonstrated the transition from weak coupling regime with 1570-fold PL intensity enhancement to strong coupling regime with Rabi splitting as large as 190 meV, which to the best of our knowledge marks the largest plasmon-exciton Rabi splitting among all TMD at room temperature. The interplay between electric field enhancement and absorption necessitates an optimal spacer thickness to achieve strong coupling, which is determined to be ~ 5 nm in this work. The formation of polariton states in NCoM with 5 nm spacer is attributed to the collaborative effect of large exciton binding energy in $MoS_2$ monolayer, ultra-small mode volume and ultra-strong optical mode confinement in NCoM. Both the experimental and theoretical studies performed here provide a new guideline for selecting suitable plasmonic nanostructures to implement strong coupling operation.

## 4. Experimental Section

*Sample Fabrication:* The NCoM plasmonic nanocavities were fabricated with bottom up self-assembly method. First, an 80-nm thick Au film was e-beam evaporated on silicon wafers with 10-nm thick titanium as adhesion film. $Al_2O_3$ dielectric spacer layer was then deposited using atomic layer deposition (ALD) at 200 ℃. Meanwhile, atomically thin $MoS_2$ flakes were grown on $SiO_2$/Si substrate by CVD, and were subsequently transferred onto $Al_2O_3$-coated gold film using Poly (methyl methacrylate) (PMMA) as mechanical protection layer during transfer. Finally, the NCoM system is formed by drop-casting a diluted silver nanocube solution (nanoComposix, 1:100) onto the prepared substrates, followed by drying with $N_2$ gas. The size distribution of the silver nanocubes is from 65 to 100 nm.

*Optical Measurements:* The scattering measurements of individual NCoMs were carried out in home-built microscope settings with xenon light source (Thorlabs) focused through 100× objective lens (NA = 0.95) in dark-field geometry. The scattered light was then collected by the same lens and directed into hyperspectral system (Cytoviva) for spectral measurement of the individual particles. The emission spectra of NCoMs were measured by a confocal Raman system (Witec) with 532 nm laser focused through 100× objective lens (NA = 0.95). The laser spot size was measured as ~ 1.5 µm. The emission was then collected by a CCD camera and a Peltier-cooled photomultiplier tube coupled to a grating spectrometer.

*Numerical Simulation:* Full wave simulations of NCoMs are performed to calculate plasmonic resonance using 3D FDTD Solutions (Lumerical Inc). The silver nanocubes are positioned above an 80 nm gold mirror coated with $Al_2O_3$ with thickness varied from 5 to 25 nm. The thickness of $MoS_2$ monolayer is set to 1 nm while the lowest size of the mesh is 0.5 nm. The edges and corners of silver nanocubes are rounded by 3 nm and 5 nm, respectively, according to the SEM images. Moreover, the nanocubes are also coated with a 5 nm insulating polyvinylpyrrolidone (PVP) layer to prevent silver from oxidation in the air. The optical properties of gold, silver, $Al_2O_3$ and PVP layer are obtained from database in the software. The optical constants of $MoS_2$ monolayer are obtained from an ellipsometer (J. A. Woollam Co.) as described in Supporting Information Note S8. To simulate electric field distribution, a plane wave (total field scattered field source) incidence is used to illuminate NCoM constructs at an angle of 0°. Three monitors (frequency domain power monitors) are placed correspondingly to obtain near field electric field maps. The mode volume is obtained using "mode volume" analysis group. From this, and quality factor calculated from the lineshape at resonance, the Purcell factor is calculated as: $F_p = \frac{3}{4\pi^2}(\frac{\lambda_0}{n})^3 \frac{Q}{V}$.

**Supporting Information**

Supporting Information is available from the Wiley Online Library or from the author.

**Acknowledgements**

The authors acknowledge financial supports from the Ministry of Education (MOE2016-T2-1-052 and MOE2017-T1-002-142) and National Research Foundation of Singapore (NRFCRP12-2013-04). ZX also acknowledge the support of the National Natural Science

Foundation of China (Grant No. 11604218). XW and BKT gratefully acknowledge funding support from Ministry of Education, Singapore (MOE2015-T2-2-043).

**Author Contributions**

ToC figure

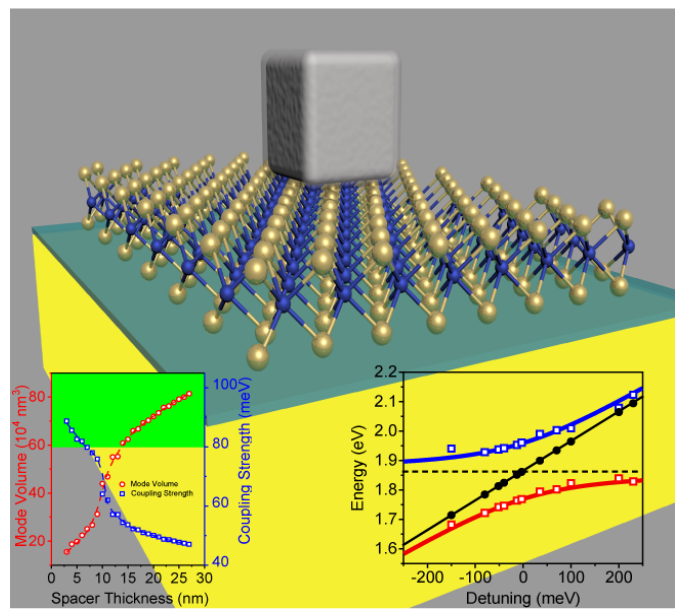

TOC description

Continuous transition from weak coupling to strong coupling is demonstrated in gap plasmonic nanocavities. The significant role of the spacer thickness on strong coupling and weak coupling from $MoS_2$ monolayer in plasmonic nanocavities has been explored experimentally and theoretically. Both the experimental and theoretical studies performed here provide a new guideline for selecting suitable plasmonic nanostructures to implement the strongest coupling operation in monolayer emitter.